\documentclass{jaa}
\usepackage[pdftex]{}
\usepackage{multicol}
\usepackage{longtable}
\usepackage{natbib}
\usepackage{graphicx}
\begin{document}\sloppy
\title{Survey of H$\alpha$ emission-line stars in the star-forming region IC~5070}
\author{Neelam Panwar\textsuperscript{1*}, Jessy Jose\textsuperscript{2}, Rishi C.\textsuperscript{1}}
\affilOne{\textsuperscript{1}Aryabhatta Research Institute of Observational SciencES, Nainital-263001, India\\}
\affilTwo{\textsuperscript{2} Indian Institute of Science Education and Research (IISER) Tirupati, Rami Reddy Nagar, Karakambadi Road, Mangalam (P.O.), Tirupati 517507, India}


\twocolumn[{

\maketitle
\corres{neelam@aries.res.in}

\msinfo{2022}{2022}


\begin{abstract}
 Actively accreting young stellar objects show H$\alpha$ emission line in their spectra. We present the results of survey for H$\alpha$ emission-line stars in the star-forming region IC~5070 taken with 2-m Himalyan {\it Chandra} Telescope. Based on the H$\alpha$ slitless spectroscopy data, we identified 131 emission-line stars in $\sim$ 0.29 square degrees area of the IC~5070 region. 
 Using Gaia Early Data Release 3, we estimated the mean proper motion and parallax of the emission-line stars. 
 We also estimated the mean distance and reddening toward the region using the emission-line stars, that are $\sim$ 833 pc and $\sim$2 mag, respectively. By examining the locations of these stars in the color-magnitude diagrams constructed using Gaia and PanSTARRS1 data, we found that a majority of the H$\alpha$ emitters are young low-mass ($<$ 1.5 $M\odot$) stars. We also compared our catalog of emission-line stars with the available young stellar catalogs and found that most of them are Class~{II}/ flat spectrum sources with the spectral type ranging from  K to M. Based on the proper motion/ parallax values and locations on the color-magnitude diagrams, about 20 emission-line stars are flagged as non-members. The  relative proper motion of the emission-line stars with respect to the ionizing source suggest the possibility of the `rocket effect' scenario in the remnant cloud (BRC~31).     
\end{abstract}
}]

\doinum{}
\artcitid{\#\#\#\#}
\volnum{000}
\year{2022}
\pgrange{1--}
\setcounter{page}{1}
\lp{1}

\section{Introduction}
Stars form out of dense molecular clouds in star-forming regions  (SFRs) and a majority of these emerge as groups/ clusters. The spatial distribution of young stars and gas in SFRs can predict the initial conditions in which these stars form \citep[e.g.,][]{goul18,kuhn20}. In addition to the above, for nearby SFRs, a sample of young stars obtained with the photometric observations can provide a complete knowledge of clustered or scattered populations \citep{getman18,herc19,panwar19,gupta21}. H$\alpha$ emission is one of the signatures used to identify young T-Tauri stars (TTSs), which arises due to the ongoing accretion process associated with these stars. Though, objective prism surveys for H$\alpha$ emission stars were the first technique to identify large populations of young stars \citep[e.g.,][]{haro53,herbig54}, with the advent of CCDs combined with grisms, much deeper surveys were possible \citep[e.g.,][]{herbig98,herbig06,pet19}. We performed a slitless H$\alpha$ survey for the search of the H$\alpha$ emission stars in the star-forming region IC 5070. 

The star forming region IC 5070 (also known as the Pelican nebula) in Cygnus has received special attention and has become the target of many observational studies for last two decades. It is associated with the H{\sc ii} region Sh2-117 or W80 \citep{wester58}. The nebula is adjacent to the North American Nebula (NGC 7000), that is now considered as a part of the same physical gas and dust cloud \citep{reip08}. The whole NGC 7000/ IC 5070 cloud complex, also known as the North American and Pelican (NAP) complex, encloses many active star-forming sites with strong obscuration attributed to the high amount of gas and dust.
The main exciting source of the H{\sc ii} region is  2MASS J20555125+4352246 (O3.5f*+O8) spectroscopic binary \citep{maiz16,com05}, lying between the NGC 7000 and IC 5070 and is embedded within in the highly extincted (A$_V$ $\sim$ 9.6 mag) dark dust cloud LDN~935. Though there is one more massive star, HD 199579 \citep[O6.5 V;][]{sota11}, it is at larger distance and quite away from the geometrical center of the H{\sc ii} region, therefore, unlikely to be a main ionizing source. 

 A number of H$\alpha$ emission-line stars and many Herbig-Haro objects have been found in the region \citep{herbig58,ogura02,bally03,bally14}. High sensitivity infrared (IR) and X-ray observations with the $\it {Spitzer}$ and $\it {Chandra}$ X-ray observatories \citep{guie09,rebu11,dami17,das21} reveal several groups/ clusters of young stars embedded in the whole NAP star forming region. Presence of several protostars, cometary nebulae, bright-rimmed cloud (BRC) 31,  and outflows driven by embedded young stars indicates  ongoing active star formation throughout the region \citep{ogura02,bally03,bally14}. Various studies yield different distance estimates for the NAP complex, ranging from about 500 pc to over 1 kpc \citep{laug07,reip08,guie09}, with recent estimate based on the Gaia DR2 data placing it at $\sim$ 800 pc \citep{bhardwaj19,zucker20}. Though located adjacent to the massive star-forming regions of Cygnus X, NAP is reported to be nearer than Cygnus X \citep{kuhn20}. Recent studies on the clustering and kinematic analyses of the NAP complex show that it contains several groups of very young stars associated with the observed molecular gas that are at different distances, spanning differences upto 150-200 pc \citep{fang20,kuhn20,das21}. 
 
 The Pelican Nebula is among one of the ideal site to study the influence of massive stars on subsequent star formation activity and evolution of the parental molecular clouds. However, the optical photometric studies of PMS stars in the region are available only for a limited sample \citep[][; and references therein]{find13,pol14,ibry18,bhardwaj19}. The $BVRI$ photometry for a sample of 17 pre-main sequence (PMS) objects was presented by \citet{pol14} in the field of the NAP complex, while \citep{froe18} found two new low-mass young stars with deep recurring eclipses in the IC~5070.

The strong radiation field and the expanding H{\sc ii} region have cleared away some of the cloud surrounding 
the IC 5070 region, making part of the young stellar content visible at optical wavelengths. Such visible young stars are distinguishable based on H$\alpha$ emission line in their spectra, which is caused by ongoing accretion processes  from a circumstellar disk through funnel flows onto the star \citep{hart16}. Since the accretion of matter on the stellar surface is episodic, the H$\alpha$ emission is thus not stationary and is therefore expected to be variable, although the timescale of variability is poorly known \citep{bhardwaj19,das21,ghosh22}. An advantage of using H$\alpha$ emission to identify young stars is that even stars with little circumstellar material, which are otherwise difficult to be identified as young through infrared photometric observations, can exhibit H$\alpha$ emission. H$\alpha$ emission-line surveys and IR surveys therefore to some extent complement each other. 

Slitless spectroscopy is a very efficient method to identify H$\alpha$ emission line sources compared to slit spectroscopy or IFU surveys, which are time consuming and expensive for telescopes. In the present work, we carried out slitless H$\alpha$ spectroscopic survey of $\sim$ 0.29 square degrees region in the IC~5070 region to search for the H$\alpha$ emission-line stars.  The organization of the paper is as follows. Section 2 present the observations, data reductions and the ancillary data obtained from published young star catalogs, Gaia early data release 3 (EDR3), and slitless spectroscopy used in this study. In Section 3, we describe the methods used to identify stellar members and characterization of the  emission-line stars  using photometric and kinematic information based on the Gaia EDR3 and PanSTARRS1 (PS1) data. Finally, Section 4 summarizes the main results of the present work.

\section{Observations \& Data Reduction}
\subsection{Optical Slitless spectroscopy}
\begin{figure*}
\centering
\includegraphics[scale = 0.9, trim = 125 0 120 0, clip]{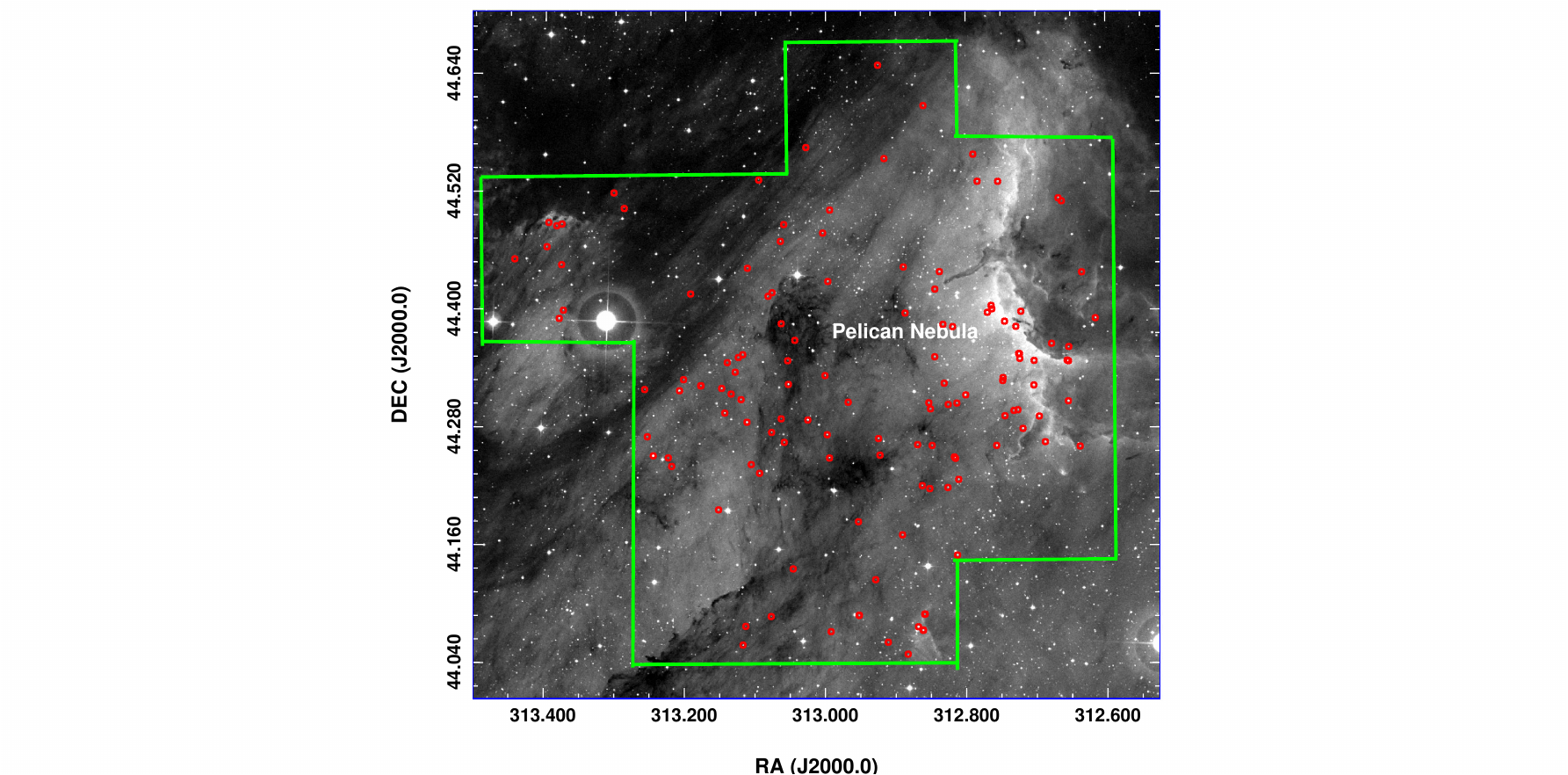}
\includegraphics[scale = 0.43, trim = 0 0 0 0, clip]{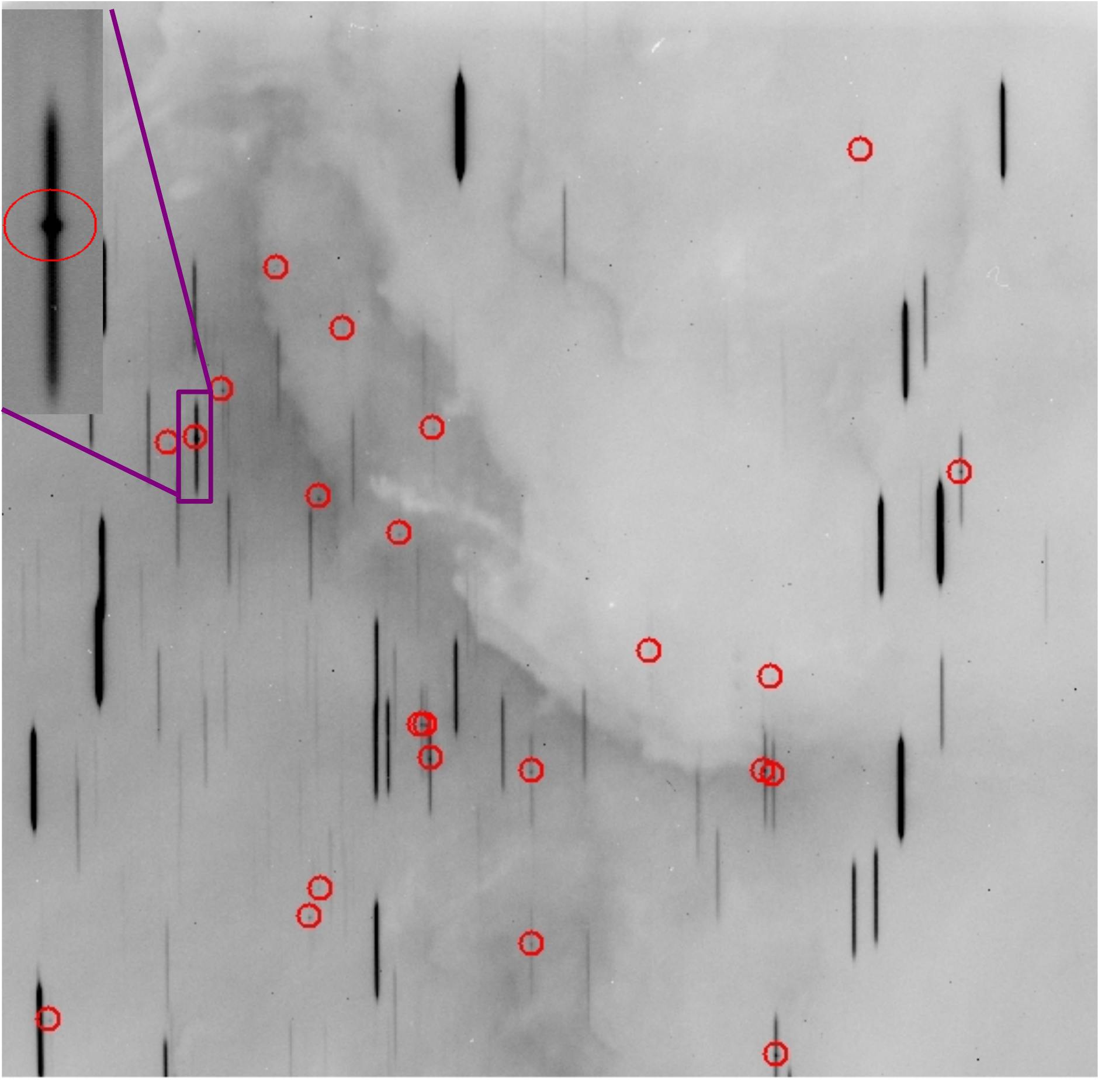}
\caption{Left Panel: DSS2R image of the IC~5070 region. Area enclosed with the green polygon represents the coverage of the present survey of H$\alpha$ emission-line stars with the HCT ($\sim$ 0.29 square degrees). The red circles represent emission-line stars identified in the present work. Right panel: Color-inverted slitless H$\alpha$ image for one of the sub-regions. Stars with enhanced H$\alpha$ emission are marked with the red circles.}
\label{fig1}
\end{figure*}
We obtained slitless observations in H$\alpha$ for the IC~5070 region (see Fig. \ref{fig1}, left panel) to detect H$\alpha$ emission line stars toward the region using the 2.0-m Himalayan {\it Chandra} Telescope (HCT) at Hanle. 
The observations were taken during the period of December-January 2012/2013. 
The details about the telescope and CCD camera can be found in \citet{chauhan09,jose2013}.
The data were acquired using a broadband H$\alpha$ filter (630-674 nm) in combination with Grism 5 (520-1030 nm) in slitless mode with the 2K $\times$ 2K CCD. The CCD has a field of view (FOV) of 10 $\times$ 10 arcmin$^2$.
The data were taken for about 11 overlapping regions, covering a total sky area ($\sim$ 0.29 square degrees), shown in Fig. \ref{fig1}. 
Images with a total integration of 1200 seconds in slitless mode were taken for each region followed by an image in H$\alpha$ filter. The image in H$\alpha$ was used to identify the coordinates of the emission-line sources. To clean the images, a number of bias and twilight flat-field frames were taken. We have used $IRAF$ software for the pre-processing of the raw images. 
The H$\alpha$ emitting sources show sharp enhancements in their spectra around 6563$\AA$ over the continuum. Based on this signature, we identified 131 sources having prominent H$\alpha$ emissions from these observations, as shown in Fig. \ref{fig1} (right panel). 
\subsection{Ancillary Data}
\subsubsection{Gaia EDR3}:
A sample of young stars obtained with the photometric observations complemented with astrometric information available with the Gaia data can be used to select the young members associated with the region. The Gaia mission has provided rich astrometric and photometric data for the stars. In the present work, we used Gaia EDR3 \citep[][]{gaia20} catalog to elucidate the kinematics and parallax of the emission-line stars in the region. 
\subsubsection{PanSTARRS data}:
The Panoramic Survey Telescope and Rapid Response System (PanSTARRS) has carried out a 3$\pi$ survey towards the northern sky ($\delta$ $>$ -30 degree). The PS1 survey was carried out mainly from at Haleakala Observatory in Hawaii in five broadband filters (g, r, i, z, y) from 2010 to 2014 \citep[see,][]{chamb16}. 
\subsubsection{Available young star catalogs}: 
The whole NAP complex is one of the most active star-forming sites and has been a target to study the clustering of young stellar objects and molecular gas. Recently, low-resolution spectroscopic survey of over about 3400 young stars in the whole NAP complex by \citet{fang20} revealed about 580 likely members. \citet{das21} identified and characterized several young stars in the NAP complex using $\it{Spitzer}$ and UKIDSS near-infrared (NIR) datasets. We have also utilized their young stellar catalogs \citep{fang20,das21} within the area covered in the present study.
\begin{figure*}
\centering
\includegraphics[scale = 0.47, trim = 0 0 0 0, clip]{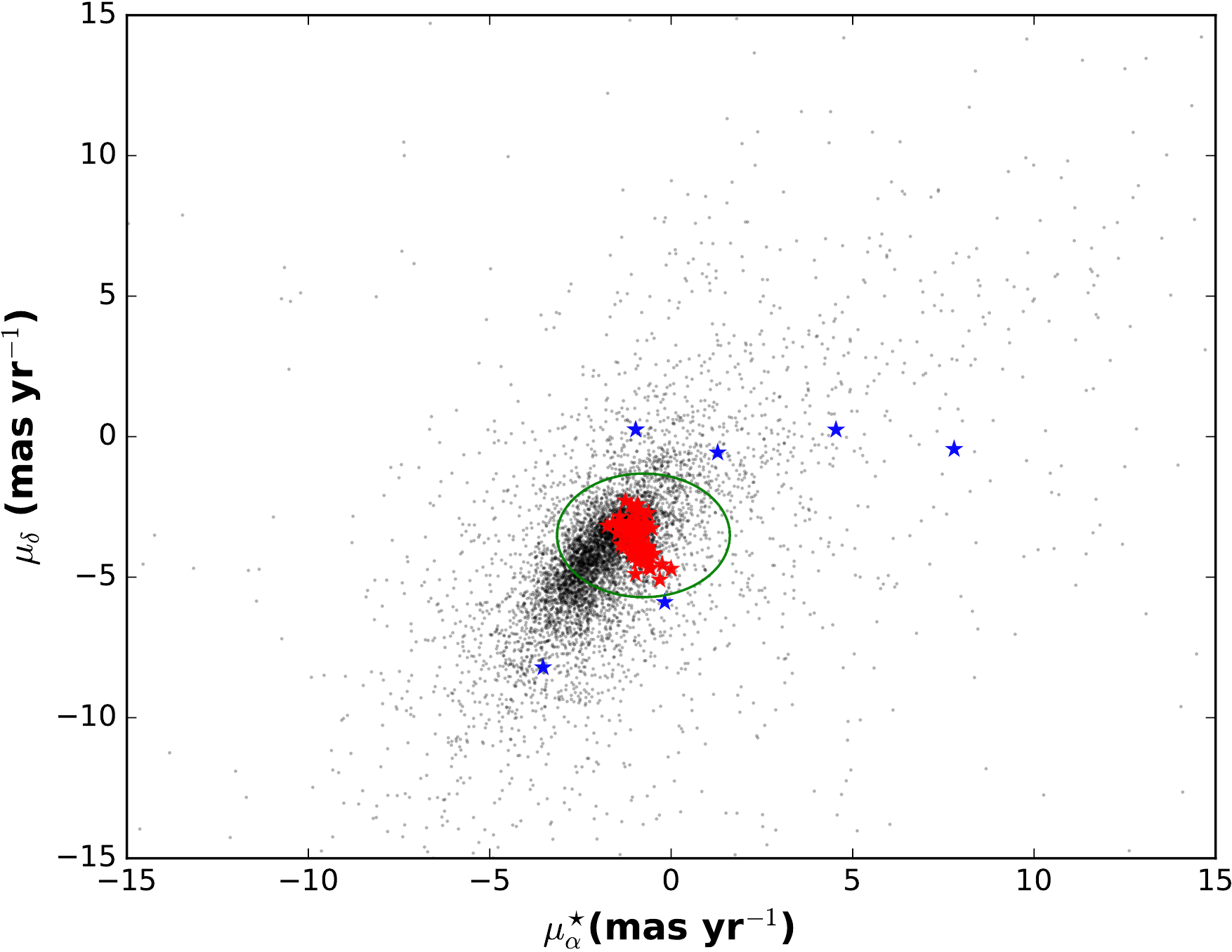}
\includegraphics[scale = 0.57, trim = 0 0 0 0, clip]{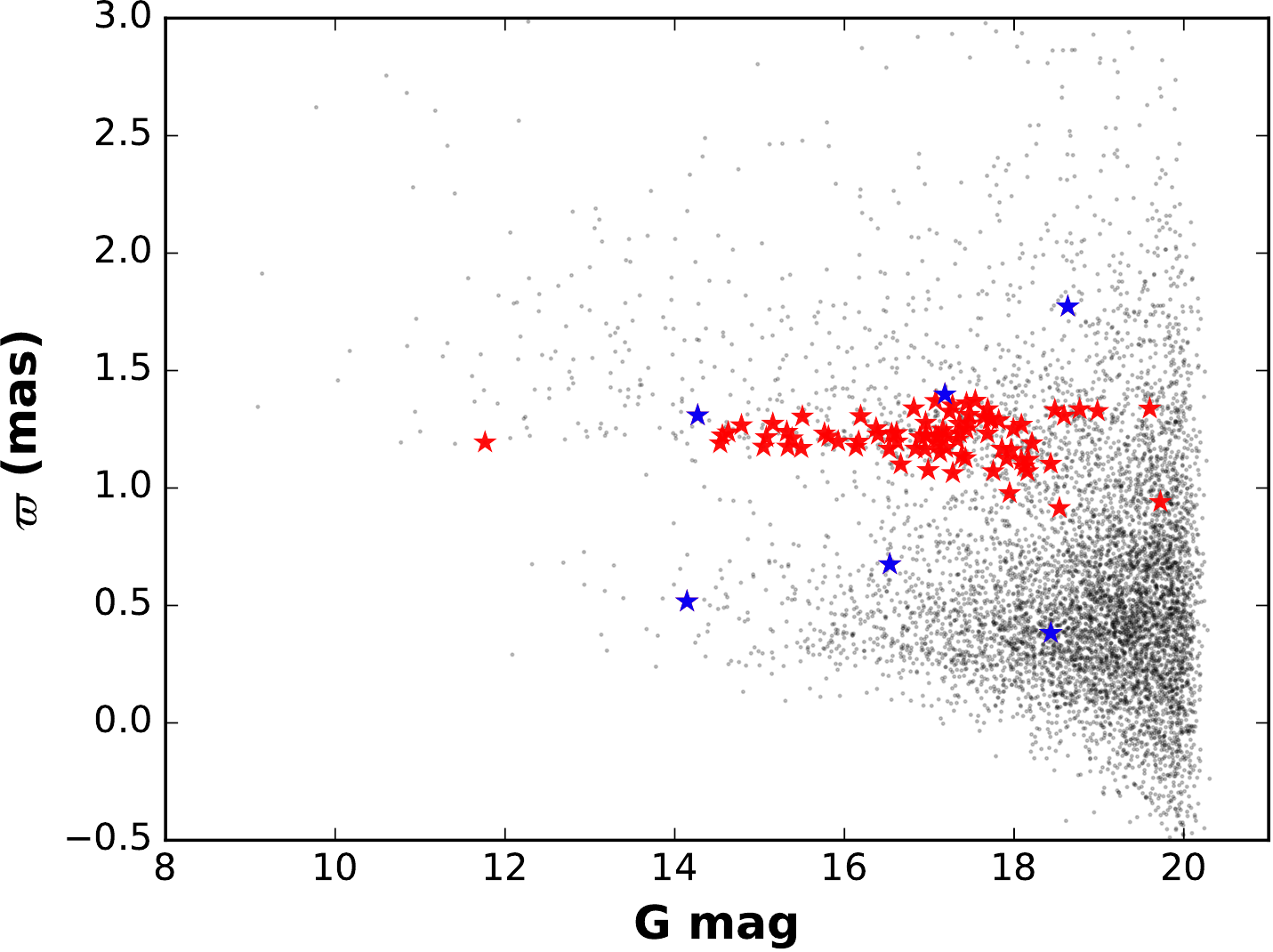}
	\caption{Vector point diagram (left panel) and  parallax vs. magnitude (right panel) for the stars in the IC~5070 region. Black dots represent the sources within the region in Fig. \ref{fig1} and red star symbols are the H$\alpha$ emission line sources detected in the present survey. The green ellipse show the three-$\sigma$ boundary around the mean proper motion values (see text).}
\label{fig2}
\end{figure*}
\section{Results \& Discussion}
\subsection{H$\alpha$ emission-line stars}
Slitless grism spectroscopy enables us to directly examine the point-like H$\alpha$ features of stars over the diffuse background emission \citep{herbig01}. We identified H$\alpha$ emission-line stars by analyzing slitless spectroscopic images of the 11 overlapping sub-regions, each of $\sim$ 10$\times$10 arcmin$^2$ area. Those stars with enhanced H$\alpha$ emission above the background are classified as emission-line stars. H$\alpha$ emission-line appear as a bright blob superposed over the continuum spectrum seen as a bright vertical line (see Fig. 1, right panel). Therefore, sources showing bright spot along the slitless spectra are designated as H$\alpha$ emission stars. In total, we identified 131 emission-line stars in the present work. The spatial distribution of these stars is  shown with the red circles in Fig. \ref{fig1}. A list of these stars is given in the Table 1.  

\subsection{Kinematics of the Identified H$\alpha$ stars}
We searched for the Gaia EDR3 counterparts of the H$\alpha$ emission-line sources within a match radius of $\sim$ 2 arcsec and obtained Gaia  measurements for $\sim$ 110 emission-line sources. 
We used the proper motion in RA ($\mu$$_\alpha$$^\star$) and proper motion in declination ($\mu_\delta$) for the young stars to generate the vector point diagram (VPD), where $\mu_\alpha^\star$ $\equiv$ ${\mu}_{\alpha}\cos(\delta)$. We restricted only those sources with proper motion uncertainty less than 0.5 mas yr$^{-1}$ and G-band magnitude uncertainty less than 0.1 mag. 
The VPD for those sources is shown in Fig. 2. The over-density of sources can be easily noticed in Fig. 2. The proper motion of the identified H$\alpha$ emission stars in the region 
peaks at $\mu_\alpha^\star$,$\mu_\delta$ $\sim$ -0.95($\pm$0.28), -3.57($\pm$0.58) mas yr$^{-1}$, which is comparable to the value found by \citet{kuhn20} for the stellar group `C', in their work, which is at the same location as that the area we considered in this study.  
An area of radius about 3-$\sigma$ (about 2.3 mas yr$^{-1}$) around the mean proper motion values of the H$\alpha$ emission-line stars is used to select the probable members (shown with green ellipse in Fig. \ref{fig2}, left panel), and the remaining sources in the VPD are considered as field stars. There were only 6 emission-line stars which seem to be non-members based on the VPD.

We also estimated the distance to the region using the parallax information of the H$\alpha$ emission stars having reliable parallax values ($\varpi$/$\sigma_\varpi$ $>$5) and are selected as proper motion members. These sources are shown with the red star symbols in Fig. \ref{fig2} (right panel). Except few outliers, the emission-line stars seem to have mean parallax values around 1.2 $mas$. After rejecting the outliers (i.e., sources with parallax values outside $\pm 3 \sigma$ of the mean),  the mean parallax value estimated for these sources 
 is $\sim$ 1.223 $\pm$ 0.077 mas. We estimate the distance after correcting the mean parallax value for the known parallax offset of $\sim$ -0.015 \citep{stas21}.  
The distance estimate for the region using Gaia data of emission line sources comes out to be $\sim$ 833 $\pm$ 52 pc, which is in agreement with that reported in previous works \citep[e.g.,][]{bhardwaj19,kuhn20,fang20}.
\subsection{Color-magnitude Diagrams: Evolutionary stages of the H$\alpha$ stars}
\begin{figure*}
\centering
\includegraphics[scale = 0.55, trim = 0 0 0 0, clip]{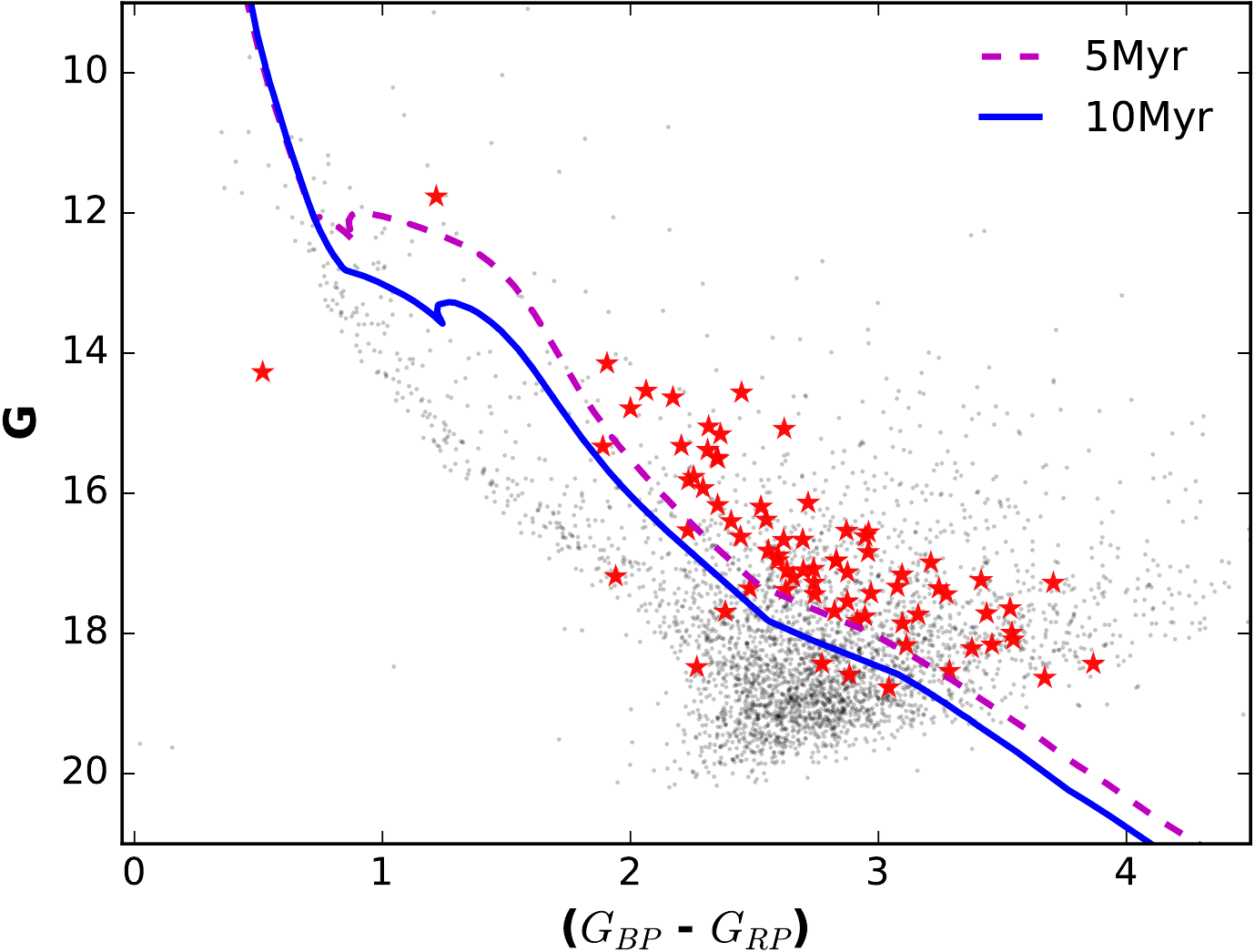}
\includegraphics[scale = 0.55, trim = 0 0 0 0, clip]{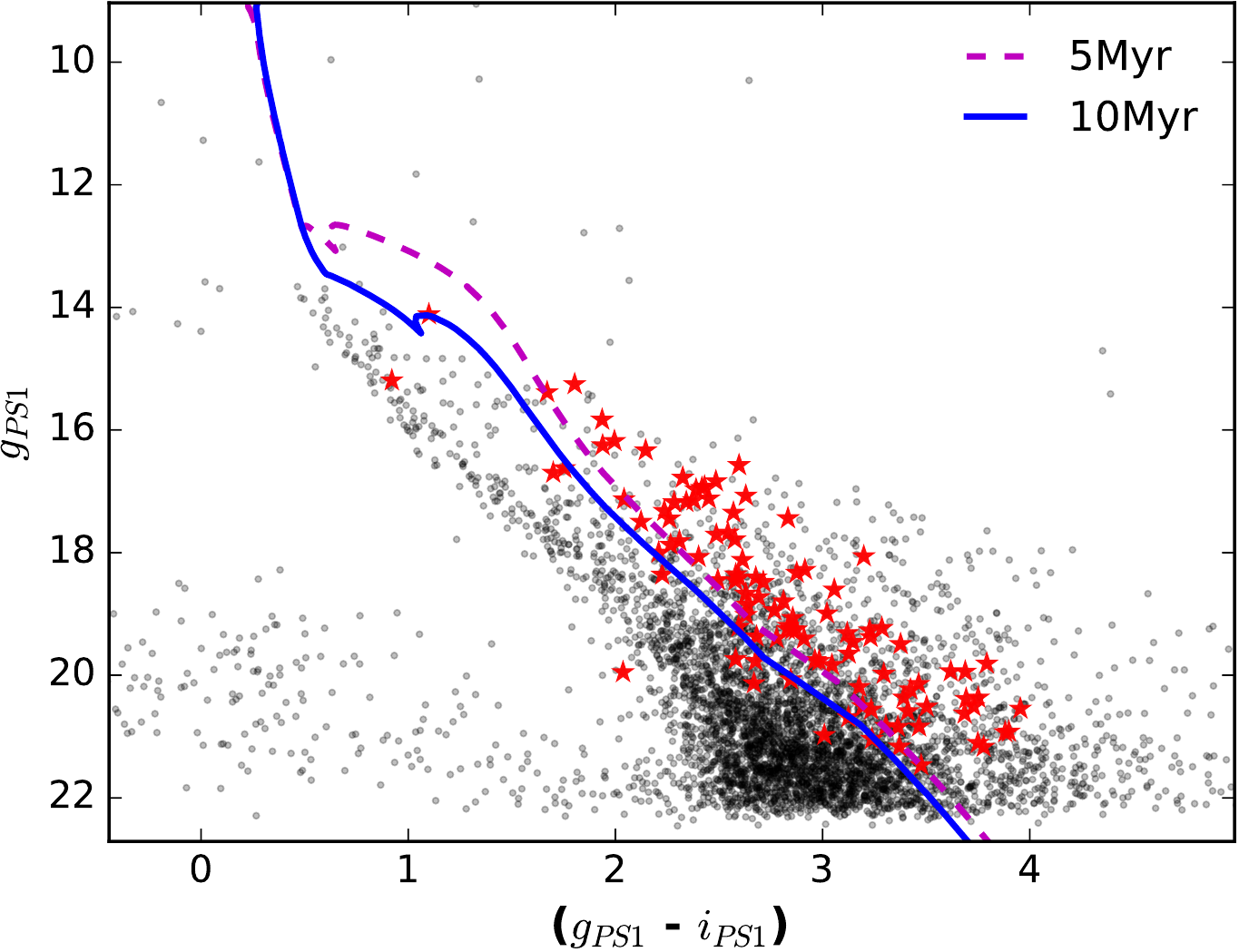}
\caption{ ($G_{BP}-G_{RP}$)/$G$ Gaia (left panel) and  ($g_{PS1}-i_{PS1}$)/$g_{PS1}$ PS1 (right panel) CMDs for the stars in the IC~5070 region. Black dots represent all the sources within the region in Fig. \ref{fig1} and red stars are the H$\alpha$ emission line sources detected in the present emission-line stars survey. The red and blue  curves are the isochrones from PARSEC models for age  5 and 10 Myr, respectively, which are corrected for the distance and average reddening of the region. }
\label{fig4}
\end{figure*}
The evolutionary stages and physical parameters of the young stars can be well constrained using the optical observations \citep{panwar2017,jose2017}. Locations of these stars on the optical color-magnitude diagrams (CMDs) can be utilized to confirm their evolutionary stages and formation mechanisms. We use optical photometry from Gaia EDR3 and PS1 surveys for the CMD analysis of the emission-line stars identified in the present work. Fig. \ref{fig4} illustrate the optical $G$/$G_{BP}-G_{RP}$)(left panel) and  $g_{PS1}$/($g_{PS1}-i_{PS1}$) (right panel) CMDs for all the stars (black dots) in the IC~5070 region. 

Star symbols represent the optical counterparts of emission-line stars in the region. Dashed and continuous curves are  5 Myr and 10 Myr isochrones for the solar metallicity taken from PARSEC \citep{pastor20}. The isochrones are shifted for the adopted distance (833 pc) and visual extinction A$_V$ $\sim$ 2 mag by adopting the extinction ratios mentioned in \citet{wang19}. In Fig. \ref{fig4}, about 90\% of the emission-line stars appear younger ($<$ 5 Myr) and a few sources are of ages $\sim$ 5-10 Myr. The recent analysis by \citet{pfal22}, shows that the median life time of low-mass stars is generally in the range of $\sim$ 5-10 Myr, which is in agreement with the age distribution of $H\alpha$ emission line sources within IC 5070. In Fig. \ref{fig4},  there are a few stars having ages more than 10 Myr. These could be either non-members of the region or the stars formed prior to the formation of ionizing source of the region, with long lived accretion  disks such as in Peter Pan Disks \citep{silver20}. Detailed spectroscopic analysis is essential to understand the nature of these sources.  The Gaia and PS1 CMDs for the emission-line sources detected in the present work, show that most of these are young members of the region with masses $<$ 1.5 M$\odot$, ie., TTSs. We flagged about 14 emission-line sources with age $>$10 Myr as non-members.

Based on the astrometric and photometric data from Gaia and PS1, we have assigned flag to the H$\alpha$ emission-line stars  cataloged in Table 1. The first digit in the column `Flag' represents the membership status of the star based on the kinematics (PMs and/or parallax), the second digit shows the membership status based on the Gaia CMD and the third digit based on the PS1 CMD. Further, `1' represents membership status `yes', `0' represents `no' and `9' represents `not confirmed'. 
Applying above criteria, we find that there are about 20 H$\alpha$ emission-line stars that are classified as non-members based on the PM/ Parallax or Gaia CMD or PS1 CMD. 
\begin{figure}
\centering
\includegraphics[scale = 0.45, trim = 0 0 0 0, clip]{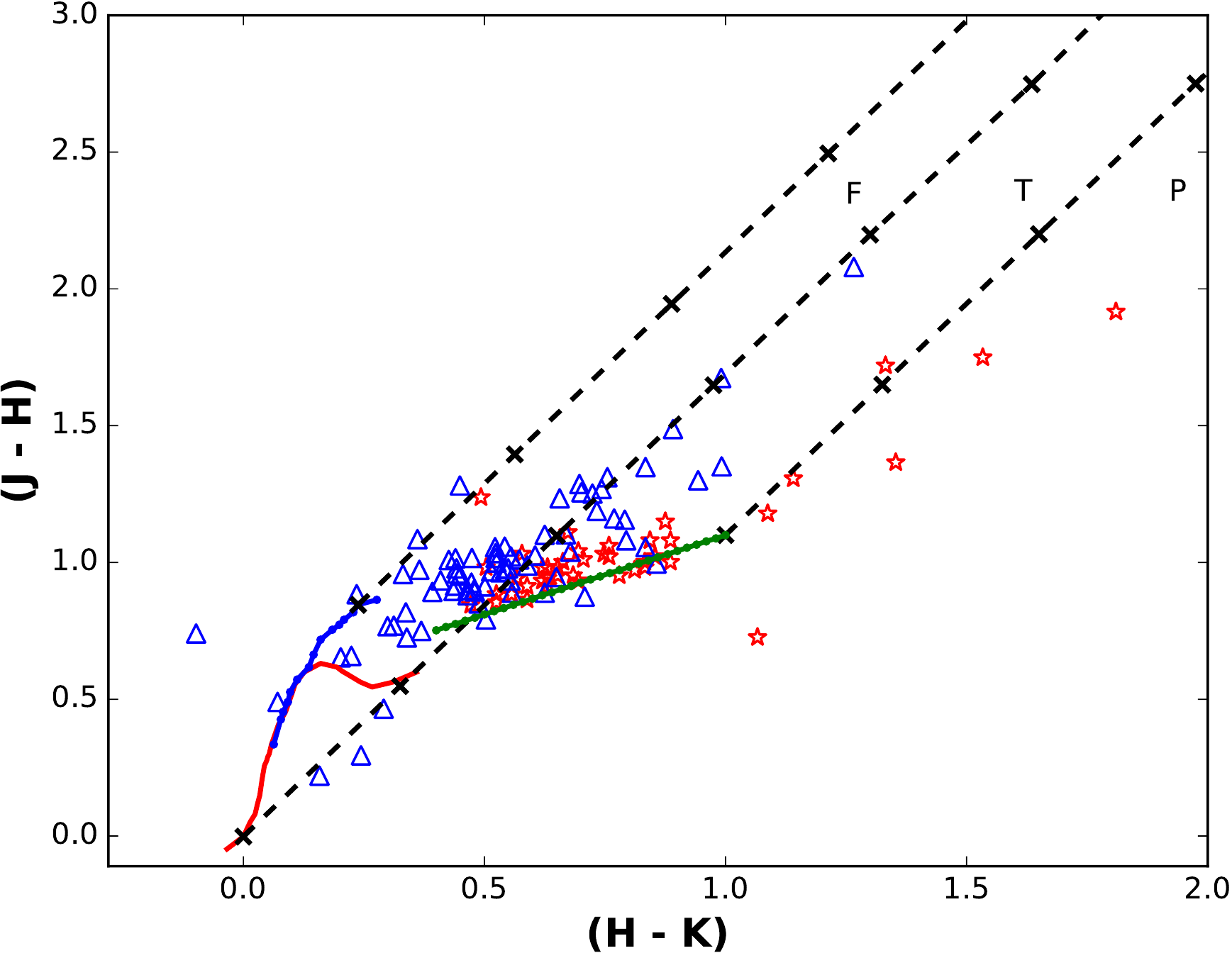}
	\caption{NIR color-color diagram for the emission-line stars in the IC~5070 region shown in Fig. \ref{fig1}. NIR counterparts of H$\alpha$ emission stars taken from YSO catalog of \citet{das21} and 2MASS data are shown with star symbols and triangles, respectively. The thick red and blue curves represent the locus of the main-sequence (MS) stars and the giants, respectively \citep{bess88}. The thin green line shows the CTTS locus \citep{mey97a}. The three parallel black dashed lines drawn from the base of the MS locus, the turning point of the MS locus, and the tip of the CTTS locus are the reddening vectors.}
\label{fig3}
\end{figure}
\subsection{Nature of the H$\alpha$ stars based on the NIR color-color diagram}
Young stellar sources exhibit excess emission in NIR wavelengths. Therefore, these sources can be identified and characterized based on their location in NIR color–color (CC) space. In Fig. \ref{fig3}, we show the 
NIR ($J$ - $H$)/($H$ – $K$) CC diagrams for the H$\alpha$ stars. To construct the CC diagram, we have used the $J$, $H$, $K$ measurements of the YSOs  from \citet{das21}, that were obtained from UKIDSS data archive. To avoid the chances of inclusion of saturated stars, we also used the 2MASS measurements for the stars which were brighter than 13.25 mag in $J$-band. The thick red curve in the CC diagram represents the main-sequence (MS) locus, and the thick blue curve is the locus of the giants \citep{bess88}. The thin green line shows the classical T Tauri star (CTTS) locus  from \citet{mey97a}. The three parallel black dashed lines drawn from the base of the MS locus, the turning point of the MS locus, and the tip of the CTTS locus are the reddening vectors. All of the magnitudes, colors, and loci of the MS, giants, and CTTS are converted to the Caltech Institute of Technology system and the extinction laws of \citet{coh81} are adopted, i.e., A$_J$ /A$_V$  = 0.265, A$_H$ /A$_V$  = 0.155, and A$_K$ /A$_V$  = 0.090. The sources in the NIR CC diagram are classified into `F', `T', and `P' regions \citep[see][]{ojha04a} that are generally considered as `MS/evolved field stars or class~{\sc iii} YSO candidates (weak-line TTSs)', `Class~{\sc ii} YSO candidates (CTTSs)' and `Class~{\sc i} YSO candidates with circumstellar envelopes', respectively. However, we note that the sources in the `T' region may be reddened early-type MS stars with excess emission in the K-band \citep{mall13} and there may be an overlap of Herbig Ae/Be stars with the sources in the `T' and `P' regions, which generally occupy the place below the CTTS locus in the NIR CC diagram \citep[for more details, see][]{hernan05}.

The NIR CC diagram of the H$\alpha$ stars clearly demonstrate that 100 H$\alpha$ emission line sources in our list are of Class~{\sc ii} and Class~{\sc iii} nature implying that about 50\% of the sources show IR excess on the NIR color-color diagram.  We also estimated the visual extinction (A$_V$) towards the region by tracing the emission-line stars to the intrinsic locus for the TTSs. The average extinction for the emission-line stars in the region is $\sim$ 2 mag.  

\subsection{Comparison with the previous YSO catalogs}
\citet{das21} identified YSOs in the whole NAP complex based on the NIR and mid-IR excess using UKIDSS and Spitzer-IRAC/ MIPS data and found many clusterings of young stars including near IC~5070. They have also classified these young stars as Class~{\sc i}, Class~{\sc ii}, flat spectrum and Class~{\sc iii} sources based on their NIR excess properties. To examine the evolutionary stages of emission-line stars, we also compared our emission-line star catalog with the YSOs catalog of \citet{das21} and found counterparts of 107 stars. Out of these 107 YSO candidates, one source is classified as Class~{\sc i}, 88 Class~{\sc ii}, 3 Class~{\sc iii}, 14 Flat spectrum sources and one is an  unclassified source by them. Hence, most of the emission-line sources identified in the present work belongs to either Class~{\sc ii} or in transition from Class~{\sc i} to Class~{\sc ii} stage (flat spectrum), indicating that H$\alpha$ emission line surveys are more efficient to detect thick disk bearing sources, i.e., actively accreting Classical T-Tauri stars. 

In order to know  the spectral type of these emission-line sources, we also cross-matched our catalog to the recent spectroscopically confirmed member and non-member catalogs of \citet{fang20}. We obtained  spectral  information corresponding to 96 emission line sources in our list. 
Most of these emission-line sources are of spectral types from K to M5. One of the emission-line star (\#43) is classified as a non-member (see their Table 6), that has the parallax (0.52 $mas$) and hence does not match with the mean parallax value of the young stars in the region. Hence, this is most probably a field star.  
\subsection{Star Formation History}
\begin{figure}
\centering
\includegraphics[scale = 0.47, trim = 0 0 0 0, clip]{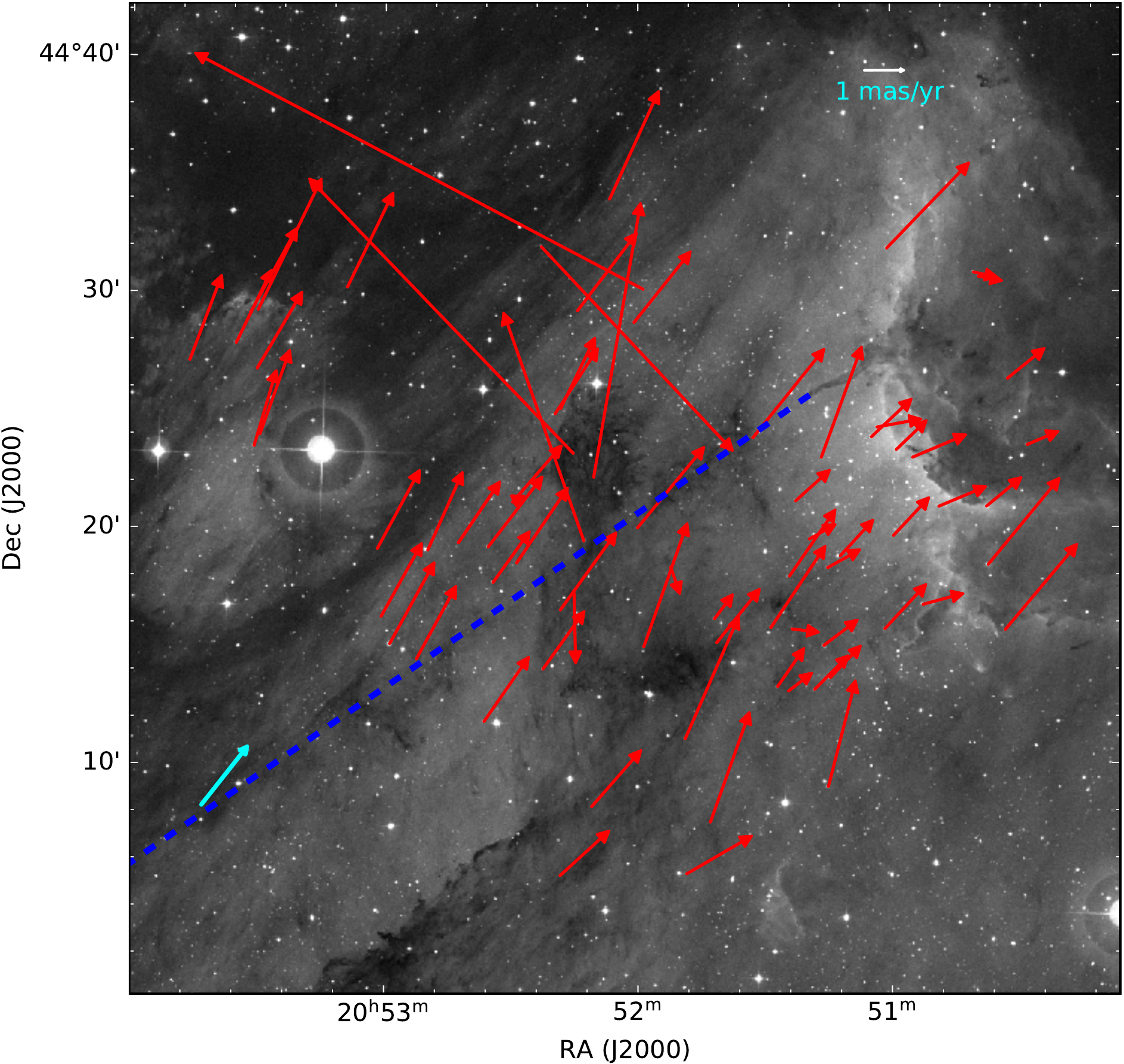}
	\caption{Proper motions of the emission line stars. The red arrows represent the relative proper motion vectors of the H$\alpha$ stars with respect to the ionizing star, obtained from Gaia EDR3. Blue dashed line indicates the directions of the main  ionizing star.}
\label{fig5}
\end{figure}
Expanding H{\sc ii} region drives an ionization/ shock fronts into the neighboring cloud, which may induce the formation of new stars due to the compression of pre-existing cores/ clumps present inside the cloud in a process known as radiation driven implosion \citep[RDI;][]{bertoldi89}. As the outer part of the molecular cloud is getting ionized by the ionization front, the ionized gas from the molecular cloud flows (photo-evaporative flow) towards the ionizing source. This results the acceleration of cloud in the opposite direction, termed as `rocket effect' \citep{oort54}. As the remnant cloud slowly moves away from the ionizing star, an aligned distribution of the young stars formed due to the RDI, can be observed between the ionizing star and remnant cloud \citep{chauhan09}. 

Small remnant molecular clouds such as BRCs, globules and elephant trunks, are found to be associated with the recent star formation activity and, in particular, BRCs are considered as ideal RDI candidates. The Pelican region consists of a bright-rimmed cloud (BRC~31) facing toward the ionizing star and contains many young stars near the bright-rim. Based on the distribution of ionized gas, bright rim, molecular gas and young stars; it is considered as one of the BRCs involved in the triggered star formation \citep{morgan09}. 

In Fig. \ref{fig1} (left panel), we notice that H$\alpha$ emitters (mostly young stars) show an aligned distribution with respect to the ionizing star. As the kinematic information of most of the H$\alpha$ emission stars in the region is available, we examined the motion of these star with respect to the main ionising source of the region, 2MASS J20555125+4352246. 
In Fig. \ref{fig5}, vectors in red color show the direction of the proper motions of the emission line stars relative to the motion of the ionizing star. The direction of the ionizing source is shown with the blue dotted line. The cyan vector shows the mean proper motion of the emission-line stars relative to the proper motion of the ionizing star. The direction of the relative proper motion of emission-line stars is away from the ionizing source and seems almost parallel with the line joining the ionizing source to the BRC~31. Our results are similar to that found by \citet{saha22}. Recently, \citet{saha22} investigated the  BRC~18 region based on the Gaia EDR3 proper motion information and suggested possibility of Rocket Effect scenario. However, we note that  here we have considered only the proper motions of the stars, and because we lack the radial velocity information of these stars, the above analysis represent only the projected motions of the stars on the sky plane.

\section{Summary}
In the present work, we report the results of our survey of emission-line stars in the star-forming region IC~5070 based on the slitless spectroscopic observations taken with the 2m-HCT. The main findings of the present work are:
\begin{itemize}
    \item We identified  131 H$\alpha$ emission-line sources in an area of $\sim$ 0.29 degree$^2$ in the IC~5070 region.
    \item The Gaia EDR3 kinematic information of the identified emission-line stars suggest that a majority of sources share a similar proper motion, except few outliers. We calculated a mean proper motion values of $\mu_\alpha^\star$,$\mu_\delta$ $\sim$ -0.95($\pm$0.28), -3.57($\pm$0.58) mas yr$^{-1}$, which are comparable to the previous works. We also used the Gaia EDR3 parallax measurements for these stars to estimate the distance of the region. Our estimated distance value of $\sim$ 833 pc is similar to that found by \citet{bhardwaj19,kuhn20,fang20}.
    \item Using NIR CC diagram for these emission-line stars, we estimated the visual extinction $A_V \sim$ 2 mag towards the region. Based on the CM diagrams constructed using Gaia and PS1 data, we find that most of the identified emission-line stars are young low-mass  (mass $<$1.5 M$\odot$, age $<$ 10 Myr), which is in agreement with the expected disk lifetime of low-mass mass stars.
    \item Based on the kinematics and photometry, we flagged the H$\alpha$ emission-line stars as members, non-members and non-confirmed sources.
    \item The relative proper motions of the emission stars with respect to the ionizing source suggests the possibility of the `rocket effect' in the BRC~31.    
 
\end{itemize}
\section*{Acknowledgements}
 We are thankful to anonymous referee for providing the constructive comments which have improved the manuscript. We thank the HTAC for time allocation and staff at the CREST \& HCT for the support during observations. NP \& RC acknowledge the financial support received through the SERB CRG2021/005876 grant and JJ acknowledges the financial support received through the DST-SERB grant  SPG/2021/003850. This work has made use of data from the European Space Agency mission {\it Gaia} (https://www.cosmos.esa.int/gaia), processed by the {\it Gaia} Data Processing and Analysis Consortium (DPAC). Funding for the DPAC
has been provided by national institutions, in particular the institutions participating in the {\it Gaia} Multilateral Agreement. This publication also makes use of data from the Two Micron All Sky Survey, which is a joint project of the University of Massachusetts and the Infrared Processing and Analysis Center/California Institute of Technology, funded by the National Aeronautics and Space Administration and the National Science Foundation. The Digitized Sky Surveys were produced at the Space Telescope Science Institute under U.S. Government grant NAG W-2166. The images of these surveys are based on photographic data obtained using the Oschin Schmidt Telescope on Palomar Mountain and the UK Schmidt Telescope. The Pan-STARRS1 Surveys were made possible through contributions by the Institute for Astronomy, the University of Hawaii, the Pan-STARRS Project Office, the Max-Planck Society and its participating institutes, the Max Planck Institute for Astronomy, Heidelberg, and the Max Planck Institute for Extraterrestrial Physics, Garching, The Johns Hopkins University, Durham University, the University of Edinburgh, the Queen’s University Belfast, the Harvard-Smithsonian Center for Astrophysics, the Las Cumbres Observatory Global Telescope Network Incorporated, the National Central University of Taiwan, the Space Telescope Science Institute, and the National Aeronautics and Space Administration under grant no. NNX08AR22G issued through the Planetary Science Division of the NASA Science Mission Directorate, the National Science Foundation grant no. AST-1238877, the University of Maryland, Eotvos Lorand University (ELTE), and the Los Alamos National Laboratory.

\clearpage
\onecolumn
\begin{longtable}{p{.4in}p{1.0in}p{1.0in}p{0.8in}|p{.4in}p{1.0in}p{1.0in}p{0.8in}}
\caption{A list of identified H$\alpha$ emission-line stars}\\
\hline
	{\textbf{S. No.}} & {\textbf{RA (J2000.0) (\textdegree)}} &{\textbf{DEC (J2000.0) (\textdegree)}}& {\textbf{Flag}}&{\textbf{S. No.}} &{\textbf{RA (J2000.0) (\textdegree)}} &{\textbf{DEC (J2000.0) (\textdegree)}}&{\textbf{Flag}}	
   \endfirsthead
\multicolumn{8}{c}
{{\bfseries \tablename\ \thetable{} -- continued from previous page}} \\
\hline 
	{\textbf{S. No.}} & {\textbf{RA (J2000.0) (\textdegree)}} &{\textbf{DEC (J2000.0) (\textdegree)}} &{\textbf{Flag}}&{\textbf{S. No.}} &{\textbf{RA (J2000.0) (\textdegree)}} &{\textbf{DEC (J2000.0) (\textdegree)}}&{\textbf{Flag}}\\ 
\hline 
\endhead

\hline \multicolumn{8}{c}{{Continued on next page}} \\ \hline
\endfoot

\hline 
\hline
\endlastfoot
\hline
     1&	312.86059& +44.60798 & 919 & 50& 312.88286& +44.04933 & 999  \\ 
     2&	312.91653& +44.55398 & 191 & 51& 312.91099& +44.06148 & 911  \\ 
     3&	313.02814& +44.56527 & 101 & 52& 312.95208& +44.08898 & 999  \\ 
     4&	312.92545& +44.64909 & 911 & 53& 312.99204& +44.07230 & 191  \\ 
     5&	313.19233& +44.41601 & 199 & 54& 313.05392& +44.34819 & 191  \\ 
     6&	313.09546& +44.53210 & 019 & 55& 313.04382& +44.36894 & 019  \\ 
     7&	313.05975& +44.48682 & 119 & 56& 313.05292& +44.32408 & 009  \\ 
     8&	313.06438& +44.46964 & 911 & 57& 313.06306& +44.28903 & 019  \\ 
     9&	313.11137& +44.44232 & 991 & 58& 313.07673& +44.27514 & 119  \\ 
    10& 313.07622& +44.41742 & 100 & 59& 313.05867& +44.26504 & 911   \\ 
    11& 313.08189& +44.41378 & 111 & 60& 313.09372& +44.23352 & 111   \\ 
    12& 313.06317& +44.38587 & 019 & 61& 313.10560& +44.24240 & 199   \\ 
    13& 313.00413& +44.47816 & 111 & 62& 313.11174& +44.28550 & 091   \\ 
    14& 312.99671& +44.42889 & 911 & 63& 313.14334& +44.29481 & 111   \\ 
    15& 312.99410& +44.50163 & 009 & 64& 313.13436& +44.31419 & 900   \\ 
    16& 312.88912& +44.44361 & 911 & 65& 313.12020& +44.30853 & 111   \\ 
    17& 312.83750& +44.43878 & 911 & 66& 313.14789& +44.31988 & 111   \\ 
    18& 312.84412& +44.42104 & 119 & 67& 313.12866& +44.33646 & 910   \\ 
    19& 312.88662& +44.39667 & 111 & 68& 313.13993& +44.34614 & 991   \\ 
    20& 312.83271& +44.38514 & 199 & 69& 313.11794& +44.35423 & 111   \\ 
    21& 312.81887& +44.38278 & 110 & 70& 313.12417& +44.35142 & 111   \\ 
    22& 313.02521& +44.28781 & 911 & 71& 313.20216& +44.32875 & 900   \\ 
    23& 312.99760& +44.27282 & 911 & 72& 313.17748& +44.32255 & 111   \\ 
    24& 312.99435& +44.24912 & 111 & 73& 313.17749& +44.32255 & 119   \\ 
    25& 313.00074& +44.33306 & 111 & 74& 313.25755& +44.31863 & 111   \\ 
    26& 312.96776& +44.30594 & 111 & 75& 313.25351& +44.27050 & 111   \\ 
    27& 312.92462& +44.26895 & 111 & 76& 313.24501& +44.25110 & 119  \\ 
    28& 312.92256& +44.25198 & 111 & 77& 313.22359& +44.24906 & 199  \\ 
    29& 312.84837& +44.26183 & 111 & 78& 313.21874& +44.24034 & 111  \\ 
    30& 312.86245& +44.22097 & 111 & 79& 313.20781& +44.31744 & 111  \\ 
    31& 312.85172& +44.21787 & 111 & 80& 312.81508& +44.24867 & 911  \\ 
    32& 312.82601& +44.21896 & 111 & 81& 312.81065& +44.22721 & 111   \\
    33& 312.81672& +44.25008 & 111 & 82& 312.81308& +44.30490 & 111   \\
    34& 312.86900& +44.26263 & 111 & 83& 312.71904& +44.27894 & 111   \\
    35& 312.85063& +44.29911 & 119 & 84& 312.74460& +44.29190 & 911   \\
    36& 312.85295& +44.30521 & 911 & 85& 312.73183& +44.29729 & 191   \\
    37& 312.82535& +44.30353 & 911 & 86& 312.72623& +44.29796 & 911   \\
    38& 312.83107& +44.32517 & 119 & 87& 312.75654& +44.26169 & 110   \\
    39& 312.84446& +44.35214 & 111 & 88& 312.69575& +44.29146 & 911   \\
    40& 312.95230& +44.08879 & 111 & 89& 312.68728& +44.26535 & 919   \\
    41& 312.86117& +44.07353 & 911 & 90& 312.63780& +44.26076 & 111   \\
    42& 312.85928& +44.08997 & 991 & 91& 312.65429& +44.30686 & 111   \\
    43& 312.81256& +44.15019 & 011 & 92& 312.80039& +44.31325 & 111   \\
    44& 312.89065& +44.17087 & 911 & 93& 313.28738& +44.50291 & 111   \\
    45& 312.95324& +44.18422 & 111 & 94& 313.30214& +44.51869 & 911   \\
    46& 312.92882& +44.12519 & 109 & 95& 313.37614& +44.48684 & 111   \\
    47& 312.85861& +44.09017 & 999 & 96& 313.38358& +44.48502 & 111   \\
    48& 312.86145& +44.07399 & 199 & 97& 313.39493& +44.48834 & 911   \\
    49& 312.86820& +44.07723 & 900 & 98& 313.39742& +44.46370 & 111   \\
    99&	313.44332& +44.45117 & 111 &116& 312.65389& +44.34765 & 911   \\
   100&	313.37647& +44.44547 & 111 &117& 312.74766& +44.32776 & 111   \\
   101&	313.37959& +44.39075 & 111 &118& 312.74719& +44.33080 & 911   \\
   102&	313.37371& +44.39910 & 111 &119& 312.70334& +44.32317 & 911   \\
   103&	312.63470& +44.43819 & 111 &120& 312.78342& +44.53075 & 199   \\
   104&	312.61548& +44.39139 & 111 &121& 312.75411& +44.53063 & 119   \\
   105&	312.76938& +44.39725 & 111 &122& 312.78947& +44.55832 & 911   \\
   106&	312.76390& +44.40431 & 111 &123& 312.66787& +44.51364 & 111   \\
   107&	312.74483& +44.38817 & 110 &124& 312.66310& +44.51061 & 100   \\
   108&	312.72875& +44.38286 & 101 &125& 313.15198& +44.19623 & 111   \\
   109&	312.72158& +44.39822 & 911 &126& 313.04592& +44.13617 & 111   \\
   110&	312.76292& +44.40078 & 191 &127& 313.07698& +44.08767 & 111   \\
   111&	312.72327& +44.35025 & 919 &128& 313.11303& +44.07760 & 911   \\
   112&	312.70292& +44.34814 & 111 &129& 313.11698& +44.05859 & 911   \\
   113&	312.67791& +44.36552 & 911 &130& 312.72374& +44.35514 & 911   \\
   114&	312.65340& +44.36214 & 999 &131& 312.72488& +44.35520 & 911   \\
   115&	312.65577& +44.34832 & 111 &   &          &           &       \\
\hline                        
\end{longtable}
\clearpage
\twocolumn 
\bibliographystyle{apj}
\bibliography{ref}


\end{document}